\documentclass[letterpaper,10pt,onecolumn]{article}

\usepackage{lmodern}

\usepackage{color}
\usepackage{amsmath,amsfonts,amssymb,mathtools}
\usepackage{graphicx}

\usepackage[small]{titlesec}

\usepackage{fullpage}
\usepackage[pass,includehead,includeheadfoot,showframe=false]{geometry}

\usepackage[mathlines]{lineno}
\modulolinenumbers[2]

\usepackage{setspace}
\newcommand{\Prob}{\mathbb{P}} 		
\newcommand{\stsp}{\mathbb{S}} 		

\usepackage[font=small]{caption}

\def\Blue#1{#1}	
\def\linelabel#1{}

\begin{document}
\begin{center}
{\Large\bf An introduction to infinite HMMs for single molecule data analysis}
\\[2ex]
\large I Sgouralis and S Press{\'e}
\end{center}

\begin{abstract}
The hidden Markov model (HMM) has been a workhorse of single molecule data analysis and is now commonly used 
as a standalone tool in time series analysis or in conjunction with other analyses methods such as tracking.
Here we provide a conceptual introduction  to an important generalization of the HMM which is poised 
to have a deep impact across Biophysics: the infinite hidden Markov model (iHMM). 
As a modeling tool, iHMMs can analyze sequential data without \emph{a priori} setting a specific number of states as required for the traditional (finite) HMM. 
While the current literature on the iHMM is primarily intended for audiences in Statistics, 
the idea is powerful and the
iHMM's breadth in applicability outside Machine Learning and Data Science
warrants a careful exposition.
Here we explain the key ideas underlying the iHMM with a special emphasis on implementation
and provide a description of a code we are making freely available.
\Blue{In a companion article, we provide an important extension of the iHMM to accommodate complications such as drift}.
\end{abstract}

\section{Introduction}

Hidden Markov models (HMMs) \cite{rabiner1986introduction,eddy2004hidden} 
provide a method for analyzing sequential time series data 
and have been a \Blue{workhorse} 
across fields including Biology \cite{yoon2009hidden,krogh1994hidden}, Physics \cite{streit1990frequency,pikrakis2006classification,dasgupta2001dual}, and Engineering \cite{fine1998hierarchical,juang1991hidden,marroquin2003hidden}.

\begin{figure}[tbp]
\begin{center}
\includegraphics[scale=0.85]{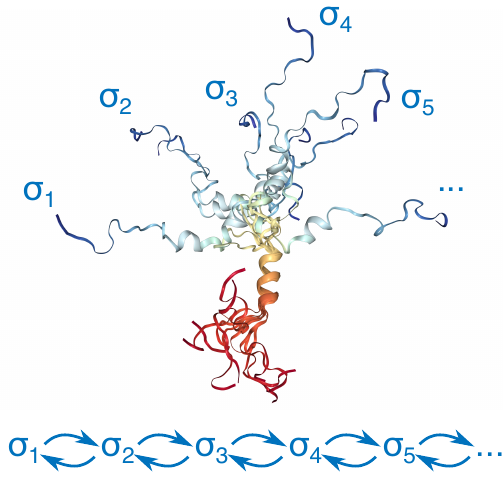}
\includegraphics[scale=0.85]{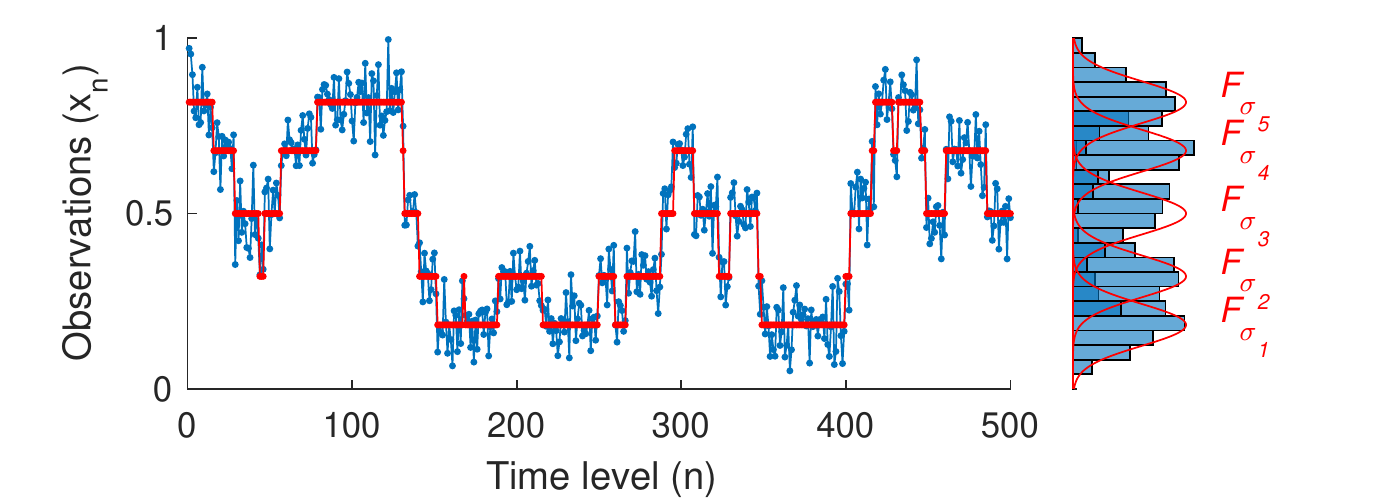}
\caption{{\bf A synthetic time trace illustrating \Blue{measurments} of a hypothetical biomolecule that undergoes conformational transitions}. 
\emph{Left:} the state space consists of conformations depicted discretely as 
$\sigma_1,\sigma_2,\dots$. 
\emph{Middle:} time series of noisy observations $x_n$ produced by the biomolecule (blue) and corresponding noiseless trace (red). Over the time \Blue{course of the measurements}, 
the biomolecule attains only conformations $\sigma_1$--$\sigma_5$, though additional conformations might be visited at subsequent times. \Blue{For sake of concreteness only, 
we label these states in order of appearance from $1$ through $5$}.
    \emph{Right:} binning the collected observations reveals ``emission distributions" $F_{\sigma_k}$ associated with each conformation. These distributions are highlighted with red lines. The centers (mean values) of the emission distributions are used to obtain the noiseless trace in the middle panel. 
The illustration on the left is created using data from Ref.~\cite{conicella2016mutations}.}
\label{fig:sys}
\end{center}
\end{figure}

The power of HMMs 
has been heavily exploited in Biophysics~\cite{mckinney2006analysis,bronson2009learning}
in the \Blue{interpretation} of single molecule experiments
such as  fluorescence resonance energy transfer (FRET) \cite{blanco2010analysis};
force spectroscopy \cite{long2013mechanical}; atomic force microscopy \cite{kruithof2009hidden}; and ion \Blue{channel} patch-clamp \cite{venkataramanan2002applying}.
The data from these disparate techniques may be analyzed using HMMs because
biomolecules or collections of biomolecules are \Blue{seen} as
visiting discrete states and the signal emitted from each state is corrupted by noise; see Fig.~\ref{fig:sys}.

While HMMs have been hugely successful, the method has encountered a fundamental limitation.  
That is, the number of \Blue{different} states visited over the course of an experiment must be specified in advance \cite{blanco2010analysis}. 
This limitation is too restrictive for complex \Blue{biomolecules}
where state numbers may often be difficult to assess {\it a priori}. It also presents a problem
when the number of states appears to vary from trace to trace as would be expected if some states are only rarely visited.
As a work-around, problem- and user-dependent \Blue{pre- or}
post-processing steps have therefore been suggested in the biophysical literature 
to winnow down the number of candidate models or eliminate problematic traces altogether.
Popular choices include model-selection tools such as 
information criteria (e.g.\ BIC, AIC~\cite{mckinney2006analysis,munro2007identification}) or maximum evidence methods (e.g.~\cite{bronson2009learning}). However, thanks to \Blue{recent} Mathematics -- Bayesian nonparametrics, first introduced in 1973~\cite{ferguson1973bayesian} -- 
an elegant solution \Blue{is available} that largely circumvents these \Blue{model selection} steps. 

In this perspectives article, we provide a description of the concepts and implementation of an important new computational tool that exploits  Bayesian nonparametrics: 
the infinite HMM (iHMM) that was heuristically described in Ref.~\cite{beal2001infinite} but only fully realized in 2012 \cite{teh2012hierarchical}.


It has recently been suggested that
Bayesian nonparametrics are poised to have a deep impact in Biophysics~\cite{hines2015primer} \Blue{and, in this field, they
have just begun to be exploited~\cite{hines2015analyzing,palla2014reversible,calderon2015inferring}\linelabel{line:hines}}.
\Blue{However, there is} -- to our knowledge -- no single resource yet available
describing the iHMM, its concepts or implementation, as would be required to bring the power of Bayesian nonparametrics to \Blue{bear}
on Biophysics and accelerate its inevitable \Blue{widespread} adoption.   
Indeed, while the iHMM tackles a conceptually simple problem, it relies on Mathematics  
whose literature is inaccessible outside a rarified community of Statisticians and Computer Scientists.   

It is for this reason that we have organized our perspective article as follows: 
Section~\ref{sec:hmm} describes the structure of the HMM in its finite (traditional) and infinite (nonparametric) realizations;
Section~\ref{sec:mcmc} presents a computational algorithm to perform inference with the iHMM that we make freely available; 
Section~\ref{sec:example} shows results from sample time traces;
and Section~\ref{sec:discussion} discusses the potential for further applications to Biophysics.

\section{Methods}
\label{sec:methods}

\subsection{Formulation of the iHMM}
\label{sec:hmm}

Here we introduce the HMM and its generalization, the iHMM. To facilitate the presentation, we initially 
describe the structure of the \Blue{space within which the biomolecule evolves and subsequently formulate the dynamics. This discussion applies to both HMMs and iHMMs.}

In the HMM framework, a system of interest is assumed to alternate successively between different states labeled $\sigma_k$
where $k$ takes integer values from $1$ to some $L$. \Blue{We use $L$ to denote the total number of states available to the system, no matter if all these states are visited or some remain unvisited during the time course of the measurements.\linelabel{line:KL_ref_1}} For instance, for an experiment on a single protein,  
the protein is the ``system"  and the ``states" are conformations such as open/closed conformations of an ion channel 
\cite{venkataramanan2002applying} \Blue{or low/high efficiency states in FRET experiments~\cite{roy2008practical}\linelabel{line:FRET_ref_1}}. \Blue{For an illustration see Fig.~\ref{fig:sys}~(left).}

In the standard HMM, we assume that the system's transitions are governed by Markovian dynamics \cite{rabiner1986introduction}. 
This means that the system jumps from a state $\sigma_k$ to a state $\sigma_j$, \Blue{for example from one FRET value to another\linelabel{line:FRET_ref_2}}, in a stochastic manner that 
depends exclusively on $\sigma_k$ and not any other state visited in the past. 
For this reason, all transitions out of state $\sigma_k$ are fully described by a probability vector 
$\tilde\pi_{\sigma_k}=(\pi_{\sigma_k\to\sigma_1},\pi_{\sigma_k\to\sigma_2},\dots)$, where $\pi_{\sigma_k\to \sigma_j}$ is the probability of departing from state $\sigma_k$ and arriving to $\sigma_j$. 

A note on our notation is appropriate here. Throughout this survey we adopt \emph{tildes} to denote vectors with components over $\stsp$, 
the system's state space $\{\sigma_1,\sigma_2,\dots\}$,
such as $\tilde\pi_{\sigma_k}$. Shortly, we will adopt \emph{bars} to denote vectors with components over time.

Once the system reaches a state $\sigma_k$, observations are emitted stochastically according to a probability distribution unique to $\sigma_k$; see Fig.~\ref{fig:sys}~(right). We call this the ``emission distribution" \Blue{$F_{\sigma_k}$}. It is often practical to model the emission distributions by a general family and use \Blue{state-specific parameters $\phi_{\sigma_k}$} to distinguish its members. 
For example, to model Gaussian emissions $x$, as in Fig.~\ref{fig:sys}, we \Blue{may} choose
\begin{linenomath*}\begin{align}\label{eq:gauss}
F_{\sigma_k}(x)=\frac{1}{\sigma_{\sigma_k}\sqrt{2\pi}}\exp\left(-\frac{(x-\mu_{\sigma_k})^2}{2\sigma_{\sigma_k}^2}\right)
\end{align}\end{linenomath*}
\Blue{where $\phi_{\sigma_k}=(\mu_{\sigma_k},\sigma_{\sigma_k})$ stands for the mean $\mu_{\sigma_k}$ and standard deviation (width) $\sigma_{\sigma_k}$ of the observations produced by state $\sigma_k$. There, for example, we define $\phi_{\sigma_1} = (\mu_{\sigma_1}, \sigma_{\sigma_1})$, $\phi_{\sigma_2} = (\mu_{\sigma_2}, \sigma_{\sigma_2})$, etc, where ($\mu_{\sigma_1}$, $\mu_{\sigma_2}, ...)$ are the state FRET efficiencies and ($\sigma_{\sigma_2}$, $\sigma_{\sigma_2}$, ...) are the corresponding spreads attributed to noise.\linelabel{line:FRET_ref_3}}

Next, we let $s_n$ denote the state of the system at the $n^{th}$ time step of the experiment and $x_n$ its corresponding observation. Thus, we label with $n$ the states of the system as it evolves through time forming a sequence $s_1\to s_2\to\dots\to s_N$, with each $s_n$ being equal to some state $\sigma_k$ chosen from $\stsp$. \Blue{For completeness, we may also assume that, before the first measurement, the system is at a default state which we denote as $s_0$.} \Blue{Again, we mention that $\stsp$ might contain states that do not appear in this sequence, i.e.\ states that remain unvisited throught the experimental time course.\linelabel{line:KL_ref_2}}

\begin{figure}[tbp]
\begin{center}
\includegraphics[scale=0.50]{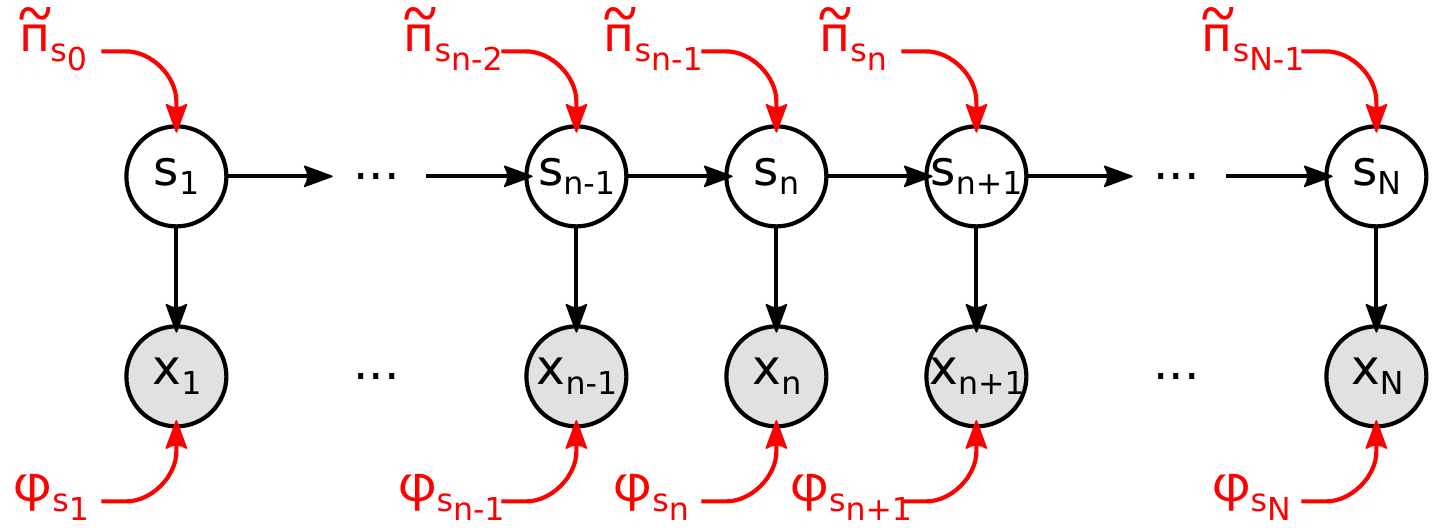}
\caption{{\bf Graphical representation of the HMM}. 
In the HMM, a biomolecule of interest transitions between unobserved states $s_n$ according to the probability vectors $\tilde\pi_{s_n}$ and generates observations $x_n$ according to the probability distributions \Blue{$F_{s_n}$ that depend on parameters $\phi_{s_n}$}. Here, following convention, the $x_n$'s are shaded to denote that these quantities are observed, while the $s_n$'s are hidden. Arrows denote the dependences among the model variables and red lines denote the model parameters.}
\label{fig:hmm}
\end{center}
\end{figure}

As is it common in the statistical literature -- and especially since we will borrow from this notation to introduce the iHMM -- 
we express the HMM compactly using the following scheme
\begin{linenomath*}\begin{align}
s_n\big|s_{n-1}&\sim Cat_\stsp(\tilde\pi_{s_{n-1}})	\label{eq:dyn_1}
\\
x_n\big|s_n&\sim F_{s_n}	\label{eq:dyn_2}
\end{align}\end{linenomath*}
and depict it schematically in Fig.~\ref{fig:hmm}. \Blue{The notation ``$\sim$'' we use above indicates that the random variables on the left side are distributed according to, and therefore sampled from, the probability distributions shown on the right side \cite{bishop2007pattern}\linelabel{line:sampled}. For example, Eq.~\eqref{eq:dyn_1} denotes a sampling from a categorical probability distribution that is supported on $\stsp$, i.e.\ $s_n$ equals a state $\sigma_k$ that is taken from $\stsp$ with probability $\pi_{s_{n-1}\to\sigma_k}$.\linelabel{line:cat}}

As can be seen from Eqs.~\eqref{eq:dyn_1} and \eqref{eq:dyn_2}, the HMM models the experimental output in a doubly stochastic manner:
i) the state of the system $s_n$ evolves stochastically within $\stsp$, as expressed by Eq.~\eqref{eq:dyn_1}; and
ii) the observations $x_n$ from each $s_n$ are also emitted stochastically, as expressed by Eq.~\eqref{eq:dyn_2}. 
In particular, for single molecule experiments, these two characteristics allow the HMM to elegantly capture: i) the seemingly random biomolecular state switching; and ii) the noise corrupting the measurements \cite{blanco2010analysis,nir2006shot}.

From the experimental measurements, we gain access only to the observations $x_n$, which take the form of a time series $\bar x=(x_1,x_2,\dots,x_N)$ over some regularly spaced time intervals $t_n$. 
In the most general case, the \textbf{goals of the HMM} are to estimate: i) the underlying state sequence $\bar s=(s_1,s_2,\dots,s_N)$ which is unobserved during the measurements; and ii) 
\Blue{the} model parameters which include 
$\phi_{\sigma_k}$ of the emission distributions and the state transition probabilities $\pi_{\sigma_k\to\sigma_j}$ associated with the states in $\stsp$. \Blue{Of course, in practice these goals are feasible only for those states that are visited during the experimental time course.\linelabel{line:KL_ref_3}}

Normally, this is the extent of the HMM. That is, one fixes $L$, i.e.\ the size of the state space $\stsp$, writes down the likelihood of observing the sequence of observations 
$\bar x=(x_1,x_2,\dots,x_N)$, and maximizes this likelihood to estimate the quantities of interest.  
Extensive literature -- that has made its way into standard textbooks, for example Ref.~\cite{bishop2007pattern} -- explain each of these steps in detail.

However, in preparation for the iHMM,
we take a Bayesian route to accomplish the goals of the HMM \cite{hines2015primer}. This is a computational overkill \Blue{for some applications} but otherwise essential to its \Blue{nonparametric} generalization. 

\Blue{Following the Bayesian paradigm}~\cite{hines2015primer,yoon2008bayesian}, we assign prior probabilities to model parameters including the emission parameters $\phi_{\sigma_k}$ and transition probability vectors $\tilde\pi_{\sigma_k}$. For instance, we may assume a prior \Blue{$\phi_{\sigma_k}\sim H$} given by a common distribution for all states. However, assigning a prior to $\tilde\pi_{\sigma_k}$ is more subtle as any choice must ensure that the predicted transitions stay within the system's state space $\stsp$. It is specifically the formulation of this prior that fundamentally distinguishes the finite from the infinite variants of the HMM \cite{teh2012hierarchical}, both of which we describe next.

In the \emph{finite variant} of the HMM \cite{rabiner1986introduction,eddy2004hidden}, 
before setting the prior on $\tilde\pi_{\sigma_k}$, we must fix $L$, the total number of states in $\stsp$ or, in the language of single molecule Biophysics, conformations
available to the biomolecule. Once $L$ is fixed, the
symmetric \emph{Dirichlet distribution} is a common choice
\begin{linenomath*}\begin{align}\label{eq:Dir}
\tilde\pi_{\sigma_k}\sim Dir_{\stsp}\left(\frac{\alpha}{L},\dots,\frac{\alpha}{L}\right).
\end{align}\end{linenomath*}
Here, $Dir_\stsp(\alpha/L,\dots,\alpha/L)$ denotes the Dirichlet distribution supported on $\stsp$ with concentration parameter $\alpha$ \Blue{(for an explicit definition of the Dirichlet distribution see supporting material)}. 
Basically, this prior asserts that lacking any information on the system's kinetics, 
\Blue{the model is likely to place on average \emph{equal} probability on any possible transition $\sigma_k\to\sigma_j$\linelabel{line:Dir}.} \Blue{In this prior, the value of $\alpha$ can be used to influence the transitions departing each state. For example, with $\alpha\gg1$ the prior tends to favor uniform transitions, while with $\alpha\ll1$ the prior tends to favor sparse transitions\linelabel{line:alpha}. In other words, the resulting sequence $\bar s$ tends to contain several different states for $\alpha\gg1$, while it tends to contain fewer for $\alpha\ll1$.}

The Dirichlet prior in Eq.~\eqref{eq:Dir} offers two key advantages: 
i) it provides a non-informative prior since no specific transitions between the $L$ states are preferentially selected; and 
ii) it \Blue{combines well with (i.e. is conjugate to)} the categorical distribution of Eq.~\eqref{eq:dyn_1}
which greatly simplifies model parameter estimation.

Although such priors help lessen the computational burden, pre-setting $L$: i) ignores the data and 
thus the arbitrary choice of $L$ may cause under- or over-fitting that, in turn, has far-reaching consequences \Blue{in estimating} state kinetics; 
and ii) does not allow the model's complexity to grow in response to newly available data \Blue{(e.g.~a rare state visited later or in another time series)}. 
Resolving issue ii) in a principled fashion \Blue{can} also help avoid cherry-picking data sets behaving closer to one's expected or preferred
value of $L$. 
It is to resolve such issues that the iHMM has been developed in the first place.

Now, in the \emph{infinite} HMM, the key difference is that $\stsp$ is assumed infinite in size \cite{beal2001infinite}. 
To avoid any confusion, we point out that this assumption is different from forcing the system to visit an infinite number of states, as it might appear at first. 
Regardless of the size of $\stsp$, the system visits only $K$ different states, where $K$ cannot exceed, for example, the number of steps $N$ in the collected time series. In practice, of course, we expect $K$ to be considerably smaller than $N$.

As we will see, with an infinite \Blue{state space}, the iHMM can recruit as many states as necessary and, in doing so, avoids underfitting by growing 
the state space as new states are visited (through the effect of the prior) but also avoids overfitting
by placing more weight on states already visited (through the effect of the likelihood). 

The generalization of the Dirichlet distribution appropriate for the infinite sized $\stsp$ is the 
\emph{Dirichlet process}
\begin{linenomath*}\begin{align}\label{eq:DP}
\tilde\pi_{\sigma_k}\sim DP_\stsp(\alpha,\tilde\beta)
\end{align}\end{linenomath*}
where $\tilde\beta$ is the \Blue{``base distribution"} of the Dirichlet process (DP) which determines \Blue{how $\tilde\pi_{\sigma_k}$ are distributed on average}. The notion of a base is only required
for the infinite DP and has no direct analog in the finite Dirichlet distribution. By contrast,
$\alpha$ is similar to the concentration of the finite HMM since it controls the sparsity of the transition probability vectors $\tilde\pi_{\sigma_k}$. 
For instance, large values of $\alpha$ make the model more likely to choose $\tilde\pi_{\sigma_k}$ that are similar for all states, while low values of $\alpha$ make the model more likely to choose $\tilde\pi_{\sigma_k}$ that differ considerably. 

\begin{figure}[tbp]
\begin{center}
\includegraphics[scale=0.50]{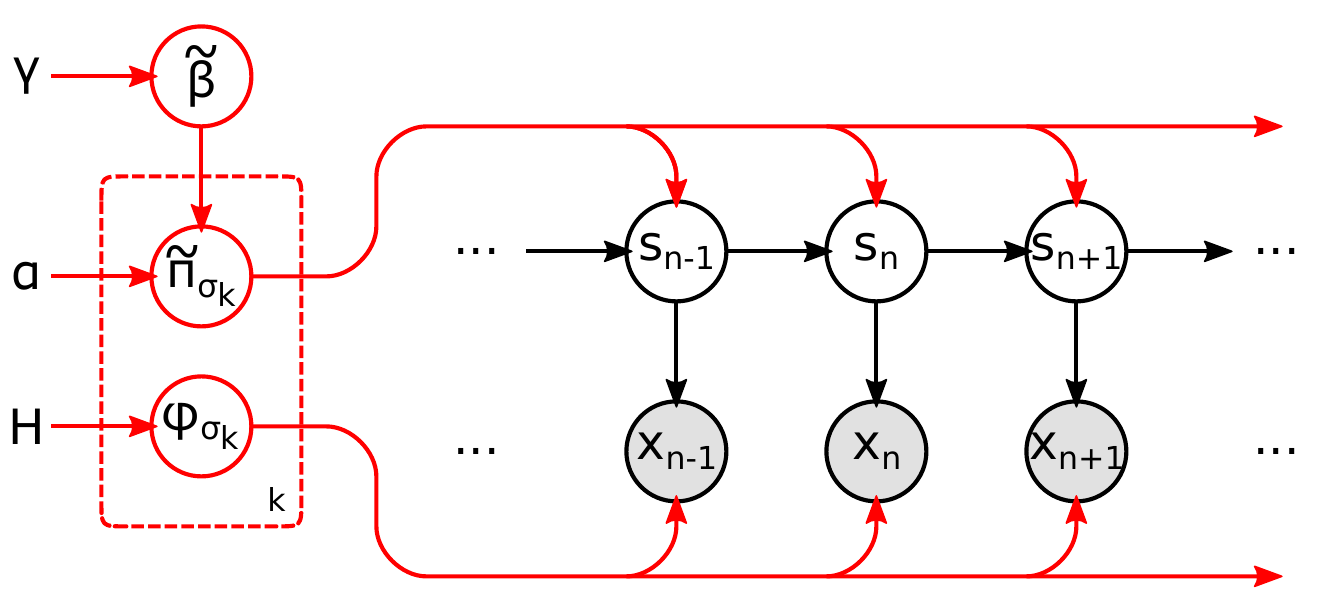}
\caption{{\bf Graphical representation of the iHMM.} The hidden Markov model that formulates the observations to be analyzed (black lines) is shown together with its priors (red lines). For completeness, we also show the concentration parameters $\alpha$, $\gamma$ and the prior probability distribution on the emission parameters $H$ that fully characterize the iHMM. The key difference from the HMM shown in Fig.~\ref{fig:hmm} is that now the model parameters $\tilde\pi_{\sigma_k}$ and $\phi_{\sigma_k}$ are treated as random variables similar to the hidden states $s_n$ and observations $x_n$. For details see the main text.}
\label{fig:ihmm}
\end{center}
\end{figure}

The DP is at the basis of much of \Blue{nonparametric Bayesian inference}\linelabel{line:DP_1}. It was only in 2012 that a hierarchical DP (HDP) was invoked in order to construct the iHMM \cite{teh2012hierarchical}.
\Blue{The basic idea is to select a random -- albeit discrete -- base distribution $\tilde\beta=(\beta_{\sigma_1},\beta_{\sigma_2},\dots)$, with members $\beta_{\sigma_k}$ that: i) are infinite in number, ii) attain non-negative values, and iii) sum to 1\linelabel{line:GEM}. In the HDP, it is common to sample such a base from a GEM (Griffiths-Engen-McCloskey) distribution realized through a
stick-breaking construction~\cite{sethuraman1994constructive, teh2012hierarchical,pitman2002poisson}}
\begin{linenomath*}\begin{align}\label{eq:GEM}
\tilde\beta&\sim GEM_\stsp(\gamma).
\end{align}\end{linenomath*}
\Blue{which ensures all these properties.}
Intuitively, the stick-breaking above is obtained as follows: we start with a stick of length 1.
We break the stick at some location $v_1$ and use the broken fraction for the probability of visiting $\sigma_1$, thus $\beta_{\sigma_1}=v_1$. Subsequently, we break the remaining stick at $v_2$ and use that new fraction for the probability of visiting $\sigma_2$, thus $\beta_{\sigma_2}=v_2(1-\beta_{\sigma_1})$, and so on. In this stick-breaking, the locations $v_k$ are sampled from a beta distribution $\mathcal{B}(1,\gamma)$, so $\gamma$ determines the relative location of the break point. Thus $\gamma$ can be used to set how rapidly $\beta_{\sigma_k}$ decreases toward zero and ultimately controls 
the number of states in $\stsp$ with an appreciable weight of being visited \Blue{at least once} during the experiment\linelabel{line:DP_2}.

In summary, the iHMM -- together with its priors illustrated in Fig.~\ref{fig:ihmm} --  takes the following form
\begin{linenomath*}\begin{align}
\tilde\beta&\sim GEM_\stsp(\gamma)	 \label{eq:ihmm_1}
\\
\tilde\pi_{\sigma_k}\big|\tilde\beta&\sim DP_\stsp(\alpha,\tilde\beta) \label{eq:ihmm_2}
\\
\phi_{\sigma_k}&\sim H \label{eq:ihmm_3}
\\
s_n\big|s_{n-1},\tilde{\tilde\pi}&\sim Cat_\stsp(\tilde\pi_{s_{n-1}}) \label{eq:ihmm_4}
\\
x_n\big|s_n,\tilde\phi&\sim F_{s_n} \label{eq:ihmm_5}
\end{align}\end{linenomath*}
where, for notational simplicity, we use $\tilde{\tilde\pi}=\{\tilde\pi_{\sigma_1},\tilde\pi_{\sigma_2},\dots\}$ and $\tilde\phi=\{\phi_{\sigma_1},\phi_{\sigma_2},\dots\}$ to gather the transition probability vectors and emission parameters associated with all states in $\stsp$. \Blue{Here, $\tilde{\tilde\pi}$ can be seen as a bi-infinite matrix whose elements are the probabilities for each possible transition $\sigma_k\to\sigma_j$\linelabel{line:trans} and it is directly analogous to the transition matrix of the finite HMM.}

Unlike the finite HMM where we need to be specific about the size of the state space, the iHMM gives us \Blue{greater flexibility through the hyperparameters $\alpha$ and $\gamma$}. For instance, larger concentrations are more appropriate for biomolecules with roughly similar transition rates and several conformations while low concentrations are more appropriate for biomolecules with dissimilar transition rates and few conformations. In general, it is possible to sum over these concentrations if we are truly ignorant of \Blue{the} properties of the biomolecule at hand~\cite{teh2012hierarchical,fox2011sticky}.

As we will see shortly, this powerful formalism allows us to infer state numbers robustly 
and, even if the number of states is apparent -- something which is rarely the case for more complex problems \cite{pirchi2011single} --
avoids the critical shortfall of cherry-picking data sets to be analyzed that appear to have similar state numbers.

\subsection{Inference on the iHMM}
\label{sec:mcmc}

Prior to describing the analysis \Blue{that iHMMs may offer}, we show how to infer quantities such as the hidden state sequence and the parameters \Blue{from the experimental traces}. A working implementation of the algorithm described in this section, equipped with a user interface, can be found in the supporting materials\linelabel{line:github}.\footnote{\Blue{This code can also be found on the authors' website as well as on GitHub.}}

To be clear, on account of the infinite size of $\stsp$, common methods used in finite state HMMs 
such as Expectation-Maximization or simply EM \cite{rabiner1986introduction} are inapplicable. 
Instead, we focus our discussion on a specialized method: the \emph{beam sampler} \cite{van2008beam}. 
Other methods are also described elsewhere \cite{teh2012hierarchical,fox2011sticky}.

The beam sampler is a special instance of the Gibbs sampler~\cite{robert2013monte}, and although it is beyond the scope of this survey, a brief description about its usage might be beneficial here. Similar to all samplers of this family, we use the beam sampler to generate (pseudo) random sequences of the model variables we are interested in inferring. Specifically, in our case these variables consist of the hidden state sequence $\bar s$ and the model parameters $\tilde\beta,\tilde{\tilde\pi},\tilde\phi$. When generating the sequences, we use Eqs.~\eqref{eq:ihmm_1}--\eqref{eq:ihmm_5} and the data $\bar x$. As a result, the generated sequences have the same statistical properties as if they were consisting of samples from the posterior probability distribution $\Prob(\bar s,\tilde\beta,\tilde{\tilde\pi},\tilde\phi|\bar x)$. So, we may use them to compute averages, confidence intervals, maximum a posteriori estimates, or simply produce histograms that resemble $\Prob(\bar s,\tilde\beta,\tilde{\tilde\pi},\tilde\phi|\bar x)$ as we show \Blue{in the next section}.

Overall, Gibbs and related samplers provide a more general approach than, for example, EM which provides only the maximum of $\Prob(\bar s,\tilde\beta,\tilde{\tilde\pi},\tilde\phi|\bar x)$. Instead, with Gibbs sampling, we fully characterize $\Prob(\bar s,\tilde\beta,\tilde{\tilde\pi},\tilde\phi|\bar x)$ over its whole domain. For an introduction to the general methodology underlying Gibbs sampling we refer the interested reader to Ref.~\cite{robert2009introducing}, while from now on we focus on the iHMM.

Suppose we have a time series of experimental observations $\bar x$. Here we describe the steps involved in generating samples (superscripted $r$) of 
our model variables $\bar s^{(r)},\tilde\beta^{(r)},\tilde{\tilde{\pi}}^{(r)},\tilde\phi^{(r)}$. \Blue{Given a computed sample $(r-1)$, the naive approach for computing the next one would be by updating each of the involved variables conditioned on the other variables and the data~\cite{robert2009introducing}}. For example, we could generate $\bar s^{(r)}$ by sampling from $\Prob(\bar s|\tilde\beta^{(r-1)},\tilde{\tilde\pi}^{(r-1)},\tilde\phi^{(r-1)},\bar x)$. Then, we could generate $\tilde\beta^{(r)}$ by sampling from $\Prob(\tilde\beta|\bar s^{(r)},\tilde{\tilde\pi}^{(r-1)},\tilde\phi^{(r-1)},\bar x$), and so on. \Blue{From the theory of Markov chain Monte Carlo sampling} -- for example Ref. \cite{robert2013monte} -- this approach is guaranteed to produce samples from the full posterior $\Prob(\bar s,\tilde\beta,\tilde{\tilde\pi},\tilde\phi|\bar x)$ as we wish. However, the infinite size of $\stsp$ makes this approach impractical since in each iteration we need to sample infinite sized $\tilde\beta,\tilde{\tilde\pi},\tilde\phi$ which is computationally \Blue{infeasible}.

\Blue{It is fundamentally because of this difficulty that we need the beam sampler which introduces a set of auxiliary variables (slicers) that in each iteration effectively truncate $\stsp$ to a finite portion~\cite{van2008beam,walker2007sampling}. In particular, for each time step $n$ in the dataset consider an auxiliary variable $u_n$ and let $\bar u=(u_1,u_2,\dots,u_N)$ gather all of them.} Since $\bar u$ consists of auxiliary variables only, we have
\begin{linenomath*}\begin{align}\label{eq:marginal}
\Prob\left(\bar s,\tilde\beta,\tilde{\tilde\pi},\tilde\phi\big|\bar x\right)=\sum_{\bar u}\Prob\left(\bar u,\bar s,\tilde\beta,\tilde{\tilde\pi},\tilde\phi\big|\bar x\right)
\end{align}\end{linenomath*}
where the sum on the right hand side is considered over all possible values $\bar u$ \Blue{may} take. Now, we can use the equality in Eq.~\eqref{eq:marginal} to sample from
$\Prob\left(\bar u,\bar s,\tilde\beta,\tilde{\tilde\pi},\tilde\phi\big|\bar x\right)$ instead from $\Prob\left(\bar s,\tilde\beta,\tilde{\tilde\pi},\tilde\phi\big|\bar x\right)$. The benefit of doing so is that, when updating $\tilde\beta,\tilde{\tilde\pi},\tilde\phi$, we have to sample conditioned also on $\bar u$.
Thus, by properly choosing the auxiliaries $u_n$, we make these updates consider only \Blue{a \emph{finite selection} of the states $\sigma_k$.}
For instance, by having $u_n$ uniformly distributed over the interval $(0,\pi_{s_{n-1}\to s_n})$, it is sufficient to consider only those $\sigma_k$ in $\stsp$ that have a probability of being visited \Blue{at least once that is greater than $u_n$.} 
Therefore, we can generate the desired samples while maintaining a finite $\stsp$ that we expand and compress dynamically according to $\bar u$. 

In summary, given $\bar s^{(r-1)}$, $\tilde\beta^{(r-1)}$, $\tilde{\tilde\pi}^{(r-1)}$, and $\tilde\phi^{(r-1)}$ the beam sampler generates a new set of samples through the following steps:
\begin{enumerate}
\item\label{st:1} Generate $\bar u^{(r)}$ by sampling from $\Prob(\bar u|\bar s^{(r-1)},\tilde{\tilde\pi}^{(r-1)})$
\item\label{st:2} Expand $\stsp$ according to $\bar u^{(r)}$
\item\label{st:3} Generate $\bar s^{(r)}$ by sampling from $\Prob(\bar s|\bar u^{(r)},\tilde{\tilde\pi}^{(r-1)},\tilde\phi^{(r-1)},\bar x)$
\item\label{st:4} Compress $\stsp$ according to $\bar s^{(r)}$
\item\label{st:5} Generate $\tilde\beta^{(r)}$ by sampling from $\Prob(\tilde\beta|\tilde\beta^{(r-1)},\bar s^{(r)})$
\item\label{st:6} Generate $\tilde{\tilde\pi}^{(r)}$ by sampling from $\Prob(\tilde{\tilde\pi}|\bar s^{(r)},\tilde\beta^{(r)})$
\item\label{st:7} Generate $\tilde\phi^{(r)}$ by sampling from $\Prob(\tilde\phi|\bar s^{(r)},\bar x)$
\end{enumerate}
where for simplicity we have dropped additional dependencies in the conditional probability distributions above.

In principle, $\tilde\beta$, $\tilde{\tilde\pi}$, and $\tilde\phi$ are infinite dimensional vectors. However, as mentioned above, the sampler requires the computation of only those components that correspond to the states which $\bar u$ allows to be visited. Therefore, in step~\ref{st:4} of the above algorithm we can safely discard those components that are absent from the state sequence $\bar s^{(r)}$. By doing so, in each time, we only need to store vectors of finite dimension, $\tilde\beta^{(r)}$, $\tilde{\tilde\pi}^{(r)}$, and $\tilde\phi^{(r)}$. Subsequently, in step~\ref{st:2} of the following iteration we can supplement $\tilde\beta^{(r)}$, $\tilde{\tilde\pi}^{(r)}$, and $\tilde\phi^{(r)}$ with states as necessary. The discarded states, since are unvisited, do not produce any of the observations in $\bar x$. Thus, they do not provide any information that could be lost. Similarly, the supplemented states are also not associated with any of the observations in $\bar x$. Thus, when we generate them, we simply have to draw samples from their priors. 
Consequently, the supplemented states neither require nor introduce extra information as it might appear at first.
In summary, expanding and compressing $\stsp$ in steps~\ref{st:2} and \ref{st:4} leaves our inference unaffected.

The supporting materials provide a pseudocode implementation of the algorithm with further details on each of the sampler's steps for the interested reader.

\section{Results}
\label{sec:example}

In this section, we demonstrate the use of the iHMM in the analysis of time series data. To benchmark the method, we use synthetic time series where the ``ground truth" is known. In a companion Research article -- that draws heavily from the formulation described in this paper -- we show the performance of the method on actual time traces that exhibit \Blue{further} complications such as drift~\cite{research}. For such time traces, we need an additional important generalization to the iHMM; the iHMM must be coupled to a continuous process. For now, our synthetic traces are meant to illustrate realistic data with noise but from which drift is otherwise absent or minimal.

Below we present the results of the analysis on two types of time traces
specifically problematic for the HMM:
\begin{itemize}
\item Dataset 1: time traces where the visited states are in high number and their emissions show significant overlap.
\item Dataset 2: time traces where some states are rarely visited and the number of states appears to change as more data become available.
\end{itemize}
Details about the generation of these datasets, \Blue{as well as more in depth analyses than those we discuss here,} can be found in the supporting materials.

\begin{figure}[tbp]
\begin{center}
\includegraphics[scale=1]{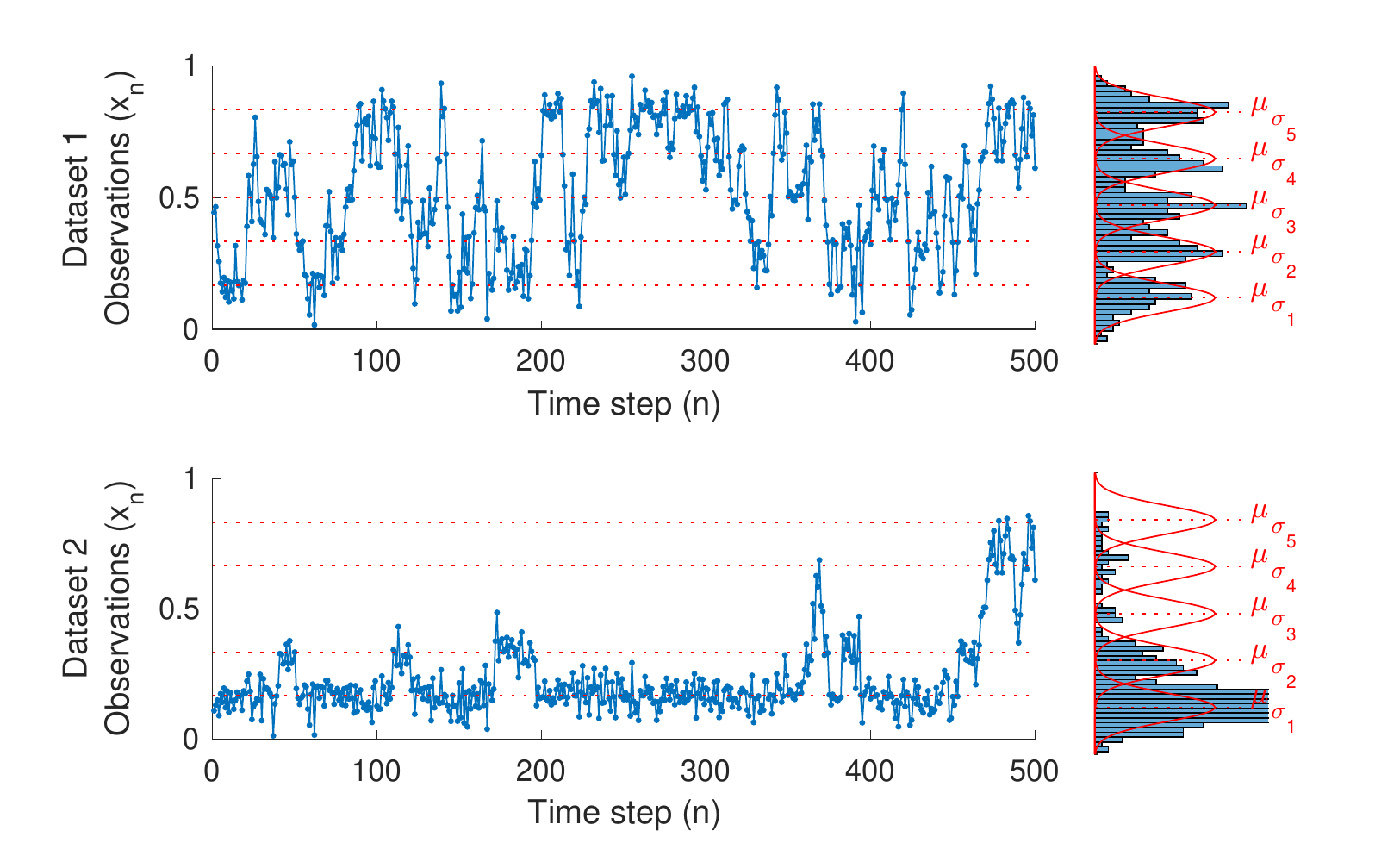}
\caption{{\bf Synthetic datasets resembling a hypothetical biomolecule undergoing transitions between discrete states that we analyzed with the iHMM}.
\emph{Left:}~time series $\bar x=(x_1,\dots,x_N)$ of noisy observations. During the measuring period the biomolecule attains 5 conformations $\sigma_1,\dots,\sigma_5$. 
The number of conformations are {\it a priori} unknown and the iHMM seeks to determine the probability over the number of states as well as their properties given the data available.
In dataset~1, the biomolecule transitions often through every state. 
By contrast, in dataset 2 transitions to some states are rare. As a result, 
all states in dataset 1 are almost equally visited throughout the experiment time course, 
while in dataset 2 higher states are visited, by chance, only toward the end of the trace. 
\emph{Right:}~corresponding emission distributions \Blue{$F_{\sigma_k}$} as obtained by simply binning the observations (blue) and by plotting the exact ones used for the simulations (red). For both datasets, the emission distributions show significant overlap. In all panels, dotted lines indicate the exact mean values $\mu_{\sigma_k}$ of the emission distributions.}
\label{fig:data}
\end{center}
\end{figure}

The datasets we analyze are shown in Fig.~\ref{fig:data}. Specifically, we used these time series to estimate the posterior distributions over:
i) the number, $K$, of different conformations \Blue{visited} by the biomolecule, and ii) \Blue{the} parameters describing the emission distributions, such as the mean value for each conformation $\mu_{\sigma_k}$, for \Blue{those} states. 
To accomplish this, we generate samples from $\Prob(\bar u,\bar s,\tilde\beta,\tilde{\tilde\pi},\tilde\phi|\bar x)$ using the beam sampler \Blue{described earlier}.

\begin{figure}[tbp]
\begin{center}
\includegraphics[scale=1]{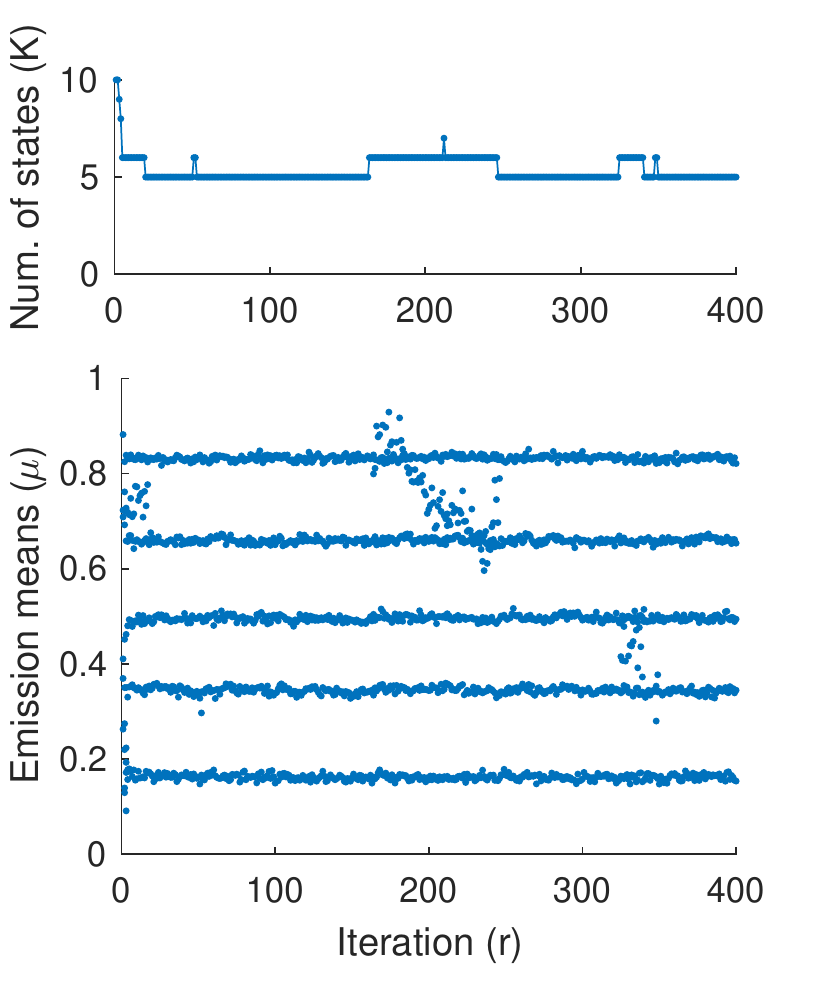}
\caption{{\bf After some iterations, the sampler used in the iHMM to analyze dataset 1 of Fig.~\ref{fig:data} eventually converges to the correct number of states}. 
The number of visited states $K^{(r)}$ (top) and the means of the emission distributions $\mu^{(r)}_{\sigma_k}$ (bottom) change throughout the sampler's iterations.
Unlike the HMM that uses a finite and fixed state space, the iHMM learns the number of available states and grows/shrinks the state-space as required by the data.}
\label{fig:iter}
\end{center}
\end{figure}

Figure~\ref{fig:iter} shows how the number of different states visited in each sampled state sequence $\bar s^{(r)}$ for dataset 1, which we denote with $K^{(r)}$,  changes through the sampler's iterations. \Blue{Again, we mention that $K$ counts only those states that are visited within the given trace.\linelabel{line:KL_ref_4}}
Initially, $K^{(1)}$ 
equals the number of states in the state space as specified in the initialization of our method. After approximately 50 iterations, $K^{(r)}$ drops to 5 which is the correct number of states attained by our hypothetical biomolecule. 

Subsequently, $K^{(r)}$ stays at 5 and occasionally adds extra states that are rapidly eliminated. 
This behavior is characteristic of our algorithm \cite{robert2013monte}. 
In the initial iterations, known as ``burn-in'', the sampler randomly explores possible choices for the model variables that eventually evolve closer to the ground truth. Once they approach the ground truth, the sampler begins generating them based on the corresponding posterior distribution $\Prob(K|\bar x)$. 
So, after burn-in, values with high $\Prob(K|\bar x)$ are sampled more often than those with low $\Prob(K|\bar x)$.

The same is true of the other variables we try to estimate. 
For example, in Fig.~\ref{fig:iter}, we also show how the sampled means of the emission distributions $\mu^{(r)}_{\sigma_k}$ change through the iterations. As with $K^{(r)}$, during burn-in, the sampler explores different choices which eventually converge to the ground truth. Afterwords, the sampler occasionally explores new emissions, for example around iterations 150--250, that quickly disappear. 

\begin{figure}[tbp]
\begin{center}
\includegraphics[scale=1]{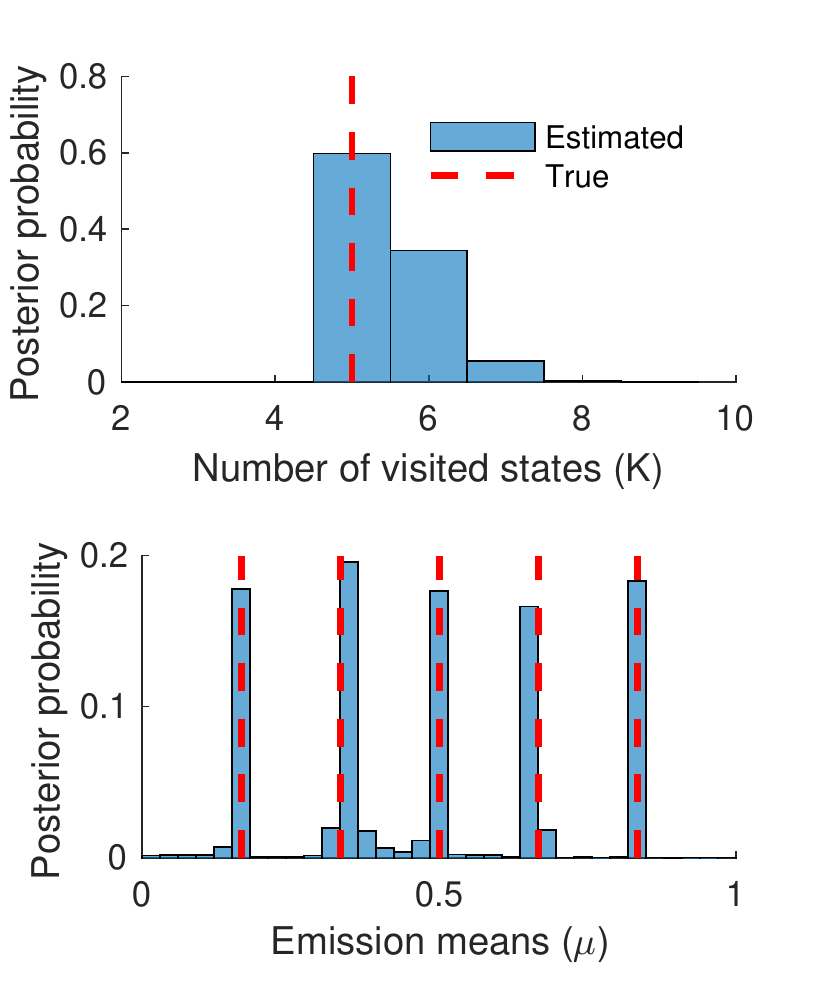}
\caption{{\bf We may use samples from the iHMM posterior probability to infer the size of the state space and the location of each state.} 
In particular, we illustrate histograms for $\Prob(K|\bar x)$ (top) and $\Prob(\mu_{\sigma_k}|\bar x)$ (bottom) using the dataset 1 of Fig.~\ref{fig:data}. On both panels dashed lines indicate the exact (ground truth) values used to produce the data in Fig.~\ref{fig:data}.}
\label{fig:hist}
\end{center}
\end{figure}

To obtain more quantitative results, we may drop the burn-in samples from the generated sequences and use the remainder of the samples to produce histograms. For example, in Fig.~\ref{fig:hist}, we show histograms of $K^{(r)}$ and $\mu_{\sigma_k}^{(r)}$. These histograms approximate the corresponding posterior probability distributions and thus provide a full characterization of the estimated variables. \Blue{In these cases, the breadth of the histograms arises because the data we analyze are limited and noisy, so the estimates are associated with some uncertainty that is reflected in the broadening of the posterior distributions.} 
 

As can be seen, dataset 1 -- while containing a large number of states which are not well separated -- can be successfully analyzed by means of the iHMM. However,
dataset 2 poses another difficulty. Namely, the biomolecule visits some states only rarely. 
In experiments, this would give rise to traces with an unequal number of states. One may naively suggest that longer traces should be collected but, due to experimental limitations \Blue{(e.g.~early photobleaching in FRET experiments\linelabel{line:FRET_ref_4}),} such long traces might not always be available. Instead, a large number of shorter traces, each containing an unknown portion of the complete \Blue{state space}, may need to be analyzed separately and the results combined~\cite{blanco2015single}. For this reason, it is crucial to treat all \Blue{traces} on an equal footing without {\it a priori} assuming a different number of states individually for each one as the HMM would require.

To illustrate how the iHMM handles such cases, we use the iHMM to analyze the data of Fig.~\ref{fig:data} (bottom trace) twice. First, using only the initial 300 time steps which visits only 2 out of the 5 states, and subsequently using the complete trace which contains all 5 states. 

\begin{figure}[tbp]
\begin{center}
\includegraphics[scale=1]{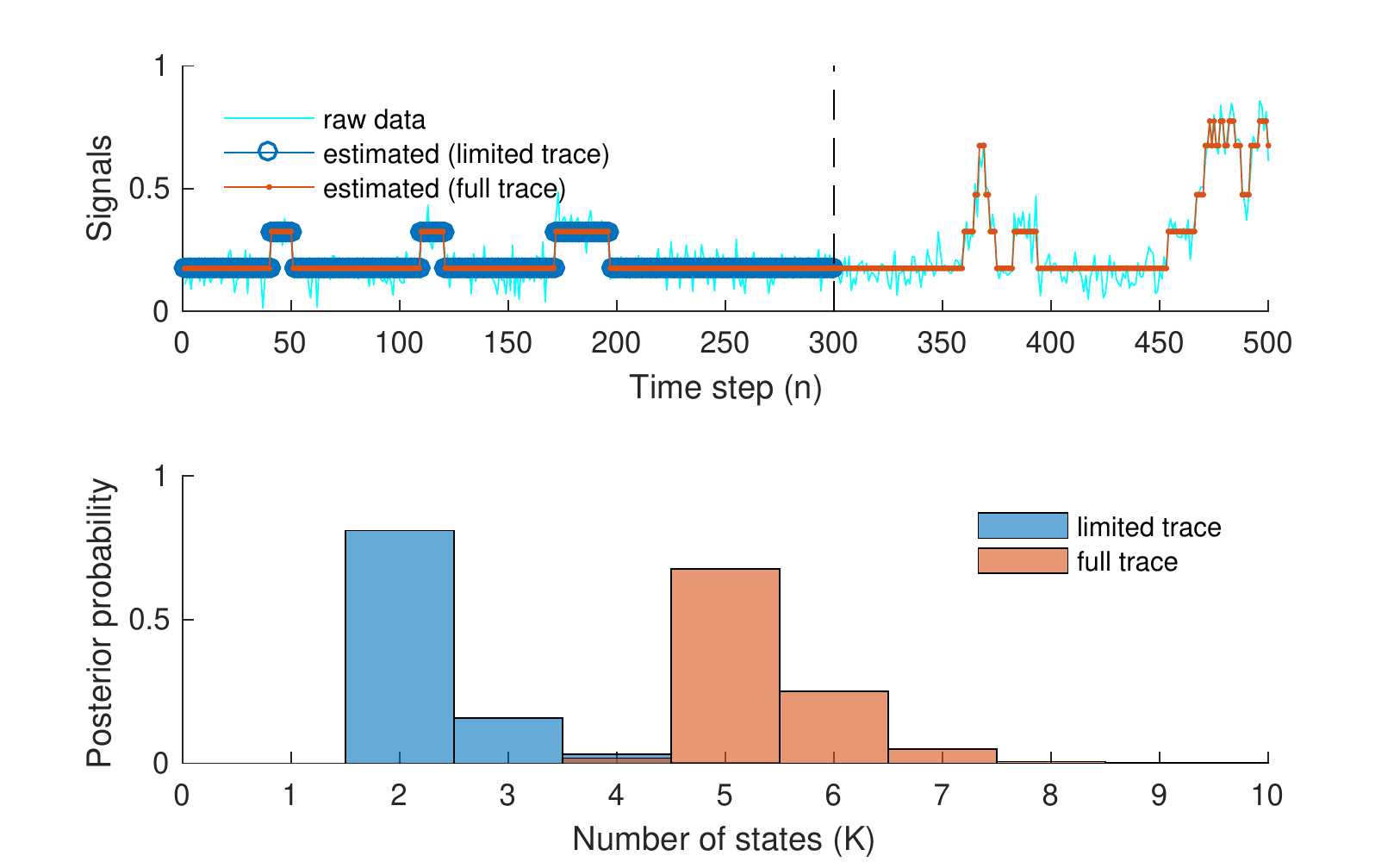}
\caption{{\bf We may use the iHMM to estimate portions of the complete state space such as those contained in different segments of dataset 2 provided in Fig.~\ref{fig:data}.} \emph{Upper:} shows estimated noiseless traces for two cases: i) using a limited segment of the full trace; and ii) using the full trace. Although only the later case allows an estimate of all 5 states, both cases provide similar estimates over those states which they mutually visit. \emph{Lower:} Corresponding estimates of the number of states contained in each trace.}
\label{fig:comp}
\end{center}
\end{figure}

Figure~\ref{fig:comp} shows the results of the analyses. Specifically, in the upper panel we show the estimated noiseless traces and in the lower panel the estimated numbers of states. As can be seen, the iHMM successfully estimates the correct number in each trace. Further, as can be seen from the denoised traces, the portion of the state space, i.e.~states $\sigma_1$ and $\sigma_2$, that is common is estimated accurately from both traces.

\section{Perspectives}
\label{sec:discussion}

In this survey we presented the infinite hidden Markov model~\cite{beal2001infinite,teh2012hierarchical}, a \Blue{relatively} recent development in Statistics which has already found diverse applicability, among other fields, in Genomics~\cite{yau2011bayesian}, Genetics \cite{xing2007hidden}, Finance~\cite{dufays2015infinite,shi2014identifying}, Tracking~\cite{kuettel2010s,fox2007hierarchical}, Machine learning~\cite{teh2010hierarchical,hu2013improved}, \Blue{and also Biophysics~\cite{hines2015analyzing,palla2014reversible,calderon2015inferring}.}

As a modeling tool, the iHMM inherits the characteristics of its predecessor -- the finite hidden Markov model -- but offers greater flexibility in the modeling and analysis of the experimental data. An important advantage is that it does not require the underlying system to have a limited state space as would its predecessor. This feature is of
particular importance to Biophysics where little information is often available to fix the number of \Blue{biomolecular} states {\it a priori}.

Despite the difficulty in prespecifying the complexity of the \Blue{state space}, 
the finite HMM is a routine choice for the analysis of single molecule time series~\cite{mckinney2006analysis,bronson2009learning,blanco2010analysis}. Nevertheless, as the experimental techniques improve and data from complex biomolecules are becoming available, the shortcomings of the finite HMM have become more apparent. To this end, methods to circumvent the limitations of HMM have been developed independently of the progress made on the same problems that have exploited Bayesian nonparametrics.

While it has previously been suggested that single molecule analysis may benefit from Bayesian nonparametrics~\cite{hines2015analyzing},
to date \Blue{only limited} applications of these methods have yet been explored. 
This is, in part, because models like the iHMM and its implementation
are scattered over various sources but also because of the important language barrier between fields
where Bayesian nonparametrics have changed the course of \Blue{data analysis} and the Physical Sciences.
It is in part for this reason that we believe methods continue to be developed to address consequences of the finiteness  
of the state space of the HMM in the Physical Sciences.

The iHMM, described here, presents great advantages in the analysis of single molecule observations. 
For one, it elegantly addresses those challenges that are presented by the HMM, namely the problem of having to select  
a number of states when the emission distributions are \Blue{unclear}. But also, to address the problem
-- currently often addressed in a user-dependent fashion -- of how to proceed when different traces present a seemingly different number of states. The key idea is that, just as the HMM performs de-noising while parametrizing transition kinetics, the iHMM
takes this logic one step further by additionally learning the number of states in a self-consistent manner. 

While greatly advantageous, the iHMM itself presents an important challenge for experimental 
data with drift, \Blue{a common feature of Biophysical time traces}~\cite{roy2008practical,blanco2010analysis}. Indeed without its explicit consideration, drift would be interpreted as the population of artifact states
by a method willing to recruit extra states to accommodate the data.
This is a key challenge that has so far not been considered anywhere. Without careful consideration, drift
would render the power of the iHMM into a key disadvantage.
While our main emphasis here has been to feature the power of the iHMM to Biophysics, our companion Research article~\cite{research} now tackles the problem of extending iHMMs to deal with drift.




\singlespacing

\bibliographystyle{unsrt}
\bibliography{ioannis_ref_ihmm}

\end{document}